# Efficient Index Maintenance Under Dynamic Genome Modification

Nitish Gupta*,1 Komal Sanjeev*,1, Tim Wall2, Carl Kingsford3, Rob Patro1

1 Department of Computer Science, Stony Brook University, Stony Brook, NY 11794-2424
2 Joint Carnegie Mellon University-University of Pittsburgh Ph.D. Program in Computational Biology
3 Department Computational Biology, School of Computer Science,
Carnegie Mellon University, 5000 Forbes Ave., Pittsburgh, PA 15213

**Abstract**

Efficient text indexing data structures have enabled large-scale genomic sequence analysis and are used to help solve problems ranging from assembly to read mapping. However, these data structures typically assume that the underlying reference text is static and will not change over the course of the queries being made. Some progress has been made in exploring how certain text indices, like the suffix array, may be updated, rather than rebuilt from scratch, when the underlying reference changes. Yet, these update operations can be complex in practice, difficult to implement, and give fairly pessimistic worst-case bounds. We present a novel data structure, SkipPatch, for maintaining a k-mer-based index over a dynamically changing genome. SkipPatch pairs a hash-based k-mer index with an indexable skip list that is used to efficiently maintain the set of edits that have been applied to the original genome. SkipPatch is practically fast, significantly outperforming the dynamic extended suffix array in terms of update and query speed.

## 1 Introduction

Static string indexing data structures such as the suffix array [11], Burrows-Wheeler transform (BWT)-based indices such as the FM-index [6], and others [18, 2] have been essential in speeding up many genomic analyses. These indices provide the ability to search the indexed string $T$ for a query string $q$ quickly. They are necessary for fast read mappers such as Bowtie [8], BWA [9], and STAR [4], for genome assembly [17], and for gene expression analysis, among many other applications.

A disadvantage of these indices is that they must be rebuilt from scratch when the underlying indexed string $T$ changes. This scenario occurs, for example, when transferring the known genetic variants of an individual (i.e., from a VCF file) to a reference genome $T$ in order to produce the individual's genome. When thousands of individuals are considered, it would be beneficial to create an index from the reference genome that could be quickly updated to be an index of the individual. Strong evidence of edits to $T$ can also occur while performing variant calling. Recently, it has been shown that iteratively updating the reference genome to account for discovered variants can significantly improve the accuracy of variant detection [7]. However, such iterative updating of the reference is time-consuming and computationally intensive when it requires rebuilding the potentially large index every time new variants are to be included in the reference sequence. Thus, practical indices that allow for fast querying in a dynamically changing string are required.

For these reasons, some dynamic string indexing data structures have been developed [15, 16, 5, 13]. Let $\Sigma$ be the alphabet and $|\Sigma|$ be its size. Let $n = |T|$. Salson et al. [15] described an algorithm that directly modifies the BWT when the text is altered. The BWT operates on cyclic shifts, and this algorithm operates based on propagating the impact created in the shifts by a single insertion. Updating the BWT has a worst case running time $\mathcal{O}(n \log n(1 + \log |\Sigma|)/ \log \log n)$. Subsequently, by exploiting the relationship between suffix arrays (SA) and BWT, Salson et al. [16] introduced an algorithm to maintain a dynamic suffix array. The worst-case update time for dynamic suffix arrays has complexity bounded by $\mathcal{O}(n \log n \log |\Sigma|)$.

---
*contributed equally to this work

A different structure, the augmented position heap, was introduced by Ehrenfeucht et al. [5]. This builds on a trie-based data structure proposed by Coffman and Eve [3], designed for storing hash tables whose keys are strings. The time to update a position heap depends on the repetitiveness of the indexed string, and is $\mathcal{O}\left(\log^3 n + b \log^2 n\right)$ in expectation, where $b$ is the length of the edit, when $T$ is chosen randomly. However, dynamic position heaps have not yet proven practical for genome-sized strings; existing implementations only support construction and search and do not support dynamic edits to the text.

We introduce a novel, practical data structure for querying a dynamically changing string. This data structure, which we call SkipPatch, allows dynamic updates to $T$ to be made in expected time $\mathcal{O}(K^2 d + \log u)$ (refer to section 2.7 for details), where $u$ is the number of edits made so far, $d$ is the time required to access a target occurrence position of a k-mer in the hash, and $K$ is the length of the k-mer. The structure supports insertions, deletions, and single nucleotide polymorphisms (SNPs). The dynamic suffix array, however, supports only insertions and deletions in its implementation. However, it can represent a SNP as an insertion/deletion pair. SkipPatch consists of a combination of a k-mer hash table and an augmented skip list that contains the edits. The data structure can be used to identify all the locations of a k-mer quickly, which is an essential operation of the seed phase for seed-and-extend read mappers. It is also capable of performing exact searches for query strings of arbitrary lengths. The data structure is primarily useful when queries in repetitive sequences do not form a significant fraction of the queries that the user intends to perform. The approach can also compute the current substring substr$(T, i, j)$ between indices $i$ and $j$ in expected time proportional to $\mathcal{O}(|j - i| + \log u)$. We also show that SkipPatch is fast to build, query, and update in practice, typically operating faster than dynamic suffix arrays, especially when $T$ is large. One of the advantages of SkipPatch is that computing the average case bounds is very straightforward. This is much more difficult in case of the Dyn SA. Thus, throughout this paper, we discuss expected case bounds for SkipPatch (which can not always be easily derived for Dyn SA). While the advantage of SkipPatch may not be reflected in its worst-case theoretical runtime bounds, our empirical results show that SkipPatch is substantially faster in terms of practical runtime.

SkipPatch is also likely to have improved cache performance compared to dynamic suffix arrays or dynamic BWT, since it avoids the scattered memory access patterns of those techniques. Evidence of this fact can be seen in the results in Section 3 detailing query speed.

For the remainder of the paper, $T$ will denote the indexed string of length $n$ over alphabet $\Sigma$.

## 2 Method

### 2.1 Overview

SkipPatch supports three different types of edit operations: SNPs (in-place substitution of a single nucleotide), insertions and deletions of arbitrary substrings. It also supports two types of queries — lookups, where occurrences of arbitrary length queries are reported, and substring extraction, where a substring of the current genome between two user-specified indices is returned.

SkipPatch consists of two main components: a k-mer index, mapping k-mers to their location on the reference, and a variant of an indexable skip list, that maintains the edits (indels). The k-mer index is a hash table that stores each length-$K$ substring of the reference as its key, and a list of the positions where it occurs on the reference genome as its value.

To perform search and edit operations, we maintain two coordinate systems. One refers to positions with respect to the initial genome, and the other refers to positions with respect to the current genome. From this point onwards, we refer to the positions in these coordinate systems as the *initial* and *current* positions. The user of the data structure (e.g. a read mapper) always refers to the genome with respect to the current positions.

## 2.2 SNPs

SNPs are a very common genomic variant. Maintenance of SNPs is performed as in-place updates. Because SNPs are so common, we implement a fast, destructive heuristic to record them: First, each k-mer that overlaps $p$ (there are up to $K$ of these) is modified. Subsequently, the reference itself is modified to record this change. Consider the reference $T = $ AGCTTTTCATTCTGA, and a SNP modifying the character at position 4 from T to G. The reference is updated to AGCTGTTCATTCTGA, and the k-mers overlapping position 4 are modified and updated in the hash. The k-mers which no longer exist (TTT) are removed from the hash table(Figure 1(a)).

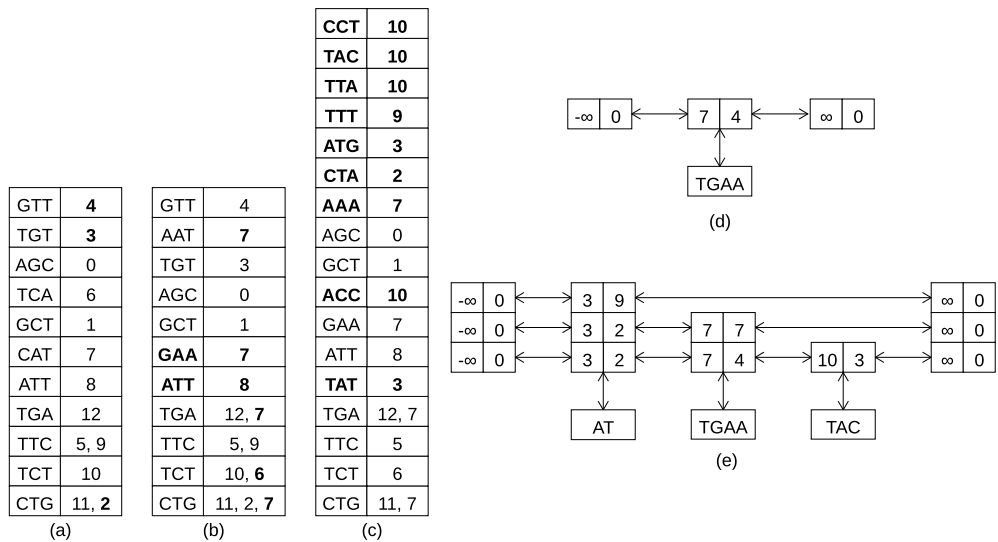

Figure 1: (a) Hash table for the reference AGCTGTTCATTCTGA and $K = 3$, after a SNP changing T at position 4 to G. The modified k-mers are in boldface. (b) Hash table for reference AGCTGTTC TGAA ATTCTGA, after inserting TGAA at position 7. (c) Hash table for reference AGCT AT GTTC TGAA ATT TAC CTGA, after inserting AT at position 3, and TAC at position 16. (d) Skip list after inserting TGAA at position 7. (e) Skip list after inserting AT at position 3, and TAC at 16.

## 2.3 Insertions and Deletions

We use an *indexable skip list* [12] to keep track of insertions and deletions. A skip list is a linked ordered sequence of elements. Querying is made faster by maintaining several linked levels of nodes above the original sequence. Each level is sparser than the previous and directly connects elements from the bottom level that are farther apart. By beginning a search at the topmost level, and at each intermediate level skipping as far as we can without exceeding the target node, skip lists (and indexable skip lists) achieve an expected $\mathcal{O}(\log n)$ search time, where $n$ is the number of elements in the skip list. To insert a new value into the skip list, a node is inserted into the bottom level of the data structure. A link directly to this element may *bubble up* to higher levels in the skip list with some probability (we set this to be 0.5 at each level). The nodes in the bottom level of the skip list store the location of the edit with respect to the *initial* reference, and are ordered by this value. We use an indexable skip list, wherein each node stores the cumulative sum of the insertion and deletion lengths between the current and the next node. Additionally, the node at the lowest level contains details of a particular edit such as the size of the insertion/deletion and the inserted text in case of an insertion. An indel changes the positions of all the k-mers occurring after the point of update. The cumulative sum of the lengths of indels before a point is used to translate *initial* positions to *current* positions, and vice versa. Since the skip list we maintain is indexable, this translation can be done in time that is logarithmic in the number of edits.

When an insertion or a deletion occurs, the original reference remains unchanged. However, it will result in addition and deletion of k-mers in the hash, and may result in creation and deletion of nodes in the skip list. SkipPatch is updated as follows:

- **Insertion**: A node is added to the skip list at the point of the insertion $p$, along with the text inserted, and its size. An insertion results in the invalidation of the k-mers crossing $p$, and addition of new k-mers into the hash at the *initial* genome position of the insertion.
- **Deletion**: A node is added to the skip list at the point where the deletion begins. This node carries a negative length representing the number of characters deleted starting from that position. A deletion results in the modification of k-mers crossing the deleted segment.

In case of nested (also known as overlapping) insertions/deletions, we edit the existing nodes as much as possible. For example, in case of inserting a segment within an existing insertion, we edit the existing skip list node to accommodate the new insertion. In case of a deletion overlapping part of an existing insertion, we remove the overlapping segment from the skip list insertion node, and if required, add a deletion node to account for the rest of the deleted segment. This manner of handling indels means that some operations can "annihilate" each other. It also results in overlapping edits having a minimal impact on the size of the skip list.

**Example.** Consider an insertion `TGAA` at position 7 in the above example (Figure 1(a)) resulting in an updated reference — `AGCTGTTC TGAA ATTCTGA`. Recording this change would require:

1. **Invalidation of k-mers crossing the point of insertion:** `TCA` and `CAT` are removed from positions 6 and 7 respectively from the hash table.
2. **Addition of new k-mers which overlap the inserted segment:** New k-mers are added — `TCT` at position 6, and `CTG`, `TGA`, `GAA`, `AAA`, `AAT` at position 7. These new k-mers are stored in the hash table with positions relative to the *initial* reference (Figure 1(b)).
3. **Insertion of a node into the skip list:** A node is inserted into the skip list with the position of insertion (7), length of insertion (4), and the inserted segment (`TGAA`) (Figure 1(d)). Now, we make two new insertions: `AT` at position 3 and `TAC` at position 16. The positions here refer to the *current* genome positions, which map back to *initial* genome positions 3 and 10 respectively. The skip list now has the structure as shown in Figure 1(e). This configuration of the skip list is obtained because the node containing position 3 has "bubbled up" twice and the node with position 7 has "bubbled up" once.

## 2.4 Substring Extraction

To extract the substring between *current* indices $i$ and $j$, we search the skip list to recover the corresponding *initial* genome positions, say $i'$ and $j'$. This is done by subtracting from $i$ and $j$ the respective cumulative sum of lengths of indels occurring before them (refer to section 2.3 for details). If there are no edits between $i'$ and $j'$, extracting the substring is a trivial substring operation on the original genome. In the case that our queried substring overlaps one or more indels, we simply begin walking through our genome and skip list in tandem, starting from $i'$ (and the *offset* in case of an insertion). We traverse the genome and skip list, splicing together the inserts and skipping deleted segments until we reach $j'$. The skip list also indicates if $i'$ and $j'$ lie on the original reference, or within an inserted segment, in which case we also know the *offset* within the inserted segment where our substring beings.

## 2.5 Reducing index space by sampling k-mers

When the reference being indexed by a data structure is large, it is often useful to perform some type of sampling to reduce the memory required for the index. This entails a trade off in which less memory is required for the index, but search becomes slower (by a constant factor). In SkipPatch, we can make such a tradeoff by sampling the k-mers that are indexed in our hash table to come from only a subset of genome positions. Specifically, when the sampling factor is set to $S$, every $S^{th}$ k-mer is hashed during index construction. The algorithms for performing edits do the same — inserting only a subset of k-mers from each insertion and making sure that k-mers deleted from particular locations are removed from the hash table. Additionally, in order to ensure no two adjacent hashed k-mers are more than distance $S$ apart on the genome, every edit is *anchored* by hashing the k-mer occurring directly following the edit location. We note that we do not allow a sampling factor, $S$, to be greater than the k-mer length ($K$). This would result in segments of the genome not being covered by any k-mer, and queries residing in such regions would not be efficiently searchable using our index.

## 2.6 Search

SkipPatch is capable of searching for query strings of arbitrary lengths. We have implemented our query procedure using an algorithm similar to the *seed-and-vote* algorithm [10] introduced by the SubRead aligner. The seed-and-vote algorithm shreds the query into smaller length strings, which we refer to as subreads. Occurrences of the query are found by determining reference positions upon which a sufficient number of subreads agree (i.e. vote). Our algorithm uses a sliding window of size $K$ to extract subreads from the query. It then finds all the locations where each of the subreads occur by querying the hash table. Each of these genome positions is then stored, along with the corresponding index (offset) of the subread within the read. These positions are then filtered and partitioned into groups of potential occurrences wherein each group adheres to the following properties:

- The list of genome positions in a group, when sorted in nondecreasing order, must have nondecreasing indices within the query.
- There are at least $(\ell - K + 1)/S$ positions in this group, where $\ell$ is the size of the query string and $S$ is the sampling factor. The search relies on the fact that no two consecutive k-mers in the hash are more than distance $S$ apart on the genome, which is ensured by the hash construction and edit algorithms. A query string of length $\ell$ can be shredded into $\ell - K + 1$ k-mers, and at least every $S^{th}$ k-mer of the query (if it exists) should appear in the hash — which gives us the $(\ell - K + 1)/S$ lower bound on the size of the group.

The genome is then read off between the starting and ending positions implied by the group, and the boundaries of this substring (the regions at the beginning and end of the string that are not covered by subreads as a result of the sampled index) are verified to ensure that the full length of the query exactly matches the genome between those positions.

When sampling is used, the minimum length of the string which can be queried is limited to $K + S - 1$. Since two consecutive hashed k-mer could be $S$ apart, in order to find an exact match, we need to ensure that at least one of the subreads of the search query is hashed, and hence searchable.

## 2.7 Running Time Analysis

- **Index build time:** The time taken to build the index is essentially the time taken to hash every k-mer of the reference, which has an amortized cost of $\mathcal{O}(Kn)$.

Updates result in k-mer modifications. Adding a new position to the list of positions of an existing k-mer takes an expected $\mathcal{O}(K)$ time, since it involves simply hashing the k-mer and appending the new position to the existing list of positions. Removing an occurrence from a k-mer's list of positions takes expected $\mathcal{O}(Kd)$ time, where $d$ is the time required to access a target position of a k-mer in the hash, and which we expect to be small for a sufficiently large $K$. Storing the k-mer positions lists, themselves, as indexable skip lists rather than simple lists could help balance the costs of insertions and deletions, and lead to faster overall updates (specifically in repetitive regions). However, we have not explored that solution in our current implementation.

- **SNP:** A SNP results in $K$ k-mer modifications; reading off these $K$ k-mers, and updating the hash table takes $\mathcal{O}\left(K^2 + K^2d + \log(u)\right)$ time in expectation. The $\log u$ factor results from traversing the skip list to read the updated genome, where $u$ is the total number of indels performed so far.
- **Insertion:** An insertion of length $\ell$ would lead to the invalidation of $K-1$ k-mers that cross the point of insertion, and an addition of $K-1+\ell$ new k-mers into to the hash table. Adding a node into the skip list indicating an insertion takes $\mathcal{O}(\log(u))$ time in expectation. Hence, an insertion of length $\ell$ takes an expected $\mathcal{O}((K-1)Kd + (K-1+\ell)K + \log(u) + s)$ time, where $s$ is the maximum size of an indel. The $\mathcal{O}(s)$ arises only in the case of a nested insertion, where the text of an existing skip list node needs to be modified.
- **Deletion:** A deletion of length $\ell$ invalidates $K+\ell-1$ k-mers that cross the deletion, and adds $K-1$ new k-mers, thus taking $\mathcal{O}((K+\ell-1)Kd + (K-1)K + \log(u) + s)$ time in expectation. The $\mathcal{O}(s)$ arises only when the deletion touches an existing skip list insertion node.
- **Substring extraction:** Reading off a segment of the updated genome requires going through the skip list. Traversing through the skip list to obtain information about the edits takes $\mathcal{O}(\log(u))$ time in expectation. Hence, the expected time to read off a segment between two *current* positions $i$ and $j$ is $\mathcal{O}(|j-i| + \log(u))$, where $|j-i|$ is the length of the substring extracted.

## 2.8 Memory Analysis

Initially, the memory consumed by SkipPatch can be attributed to the genome itself, and the k-mer hash table. The index construction hashes every k-mer of the genome, and stores it along with a list of positions at which it occurs. When a sampling factor of $S$ is used, the memory consumption is reduced by a factor of approximately $S$, which follows from the fact that the hash table constitutes the majority of the memory usage of SkipPatch. The skip list is initialized at this point, but it is empty, and therefore occupies negligible space.

As edits are performed on the reference, one expects the overall memory usage of SkipPatch to increase. The skip list occupies space of proportional to the number of edits performed on the genome, and the size of these edits if they are insertions. The insertion itself must be stored in the skip list. Considering a total of $m$ indels, and assuming that the length of each is bounded by $s$; then the memory required by the skip list will increase by $\mathcal{O}(sm)$ bytes. The other source of increasing memory usage is the modifications that must be made to the hash table as new insertions are encountered, or as SNP updates are applied. Consider an insertion of length $\ell$. This results in the modification of the $K-1$ k-mers that overlap the new insertion. Additionally, $\ell$ k-mer positions are added to the hash table (and the new k-mers that are not already present are added as well).

| Method | *E. coli* (4.6 MB) | *D. mel* (142.6 MB) | *T. nig* (288.3 MB) | *H. sap* (2.98 GB) |
| --- | --- | --- | --- | --- |
| Dyn SA | 26 | 1710 | 4622 | 39821 |
| SkipPatch | 6.4 | 191 | 429 | 207 |

Table 1: Time (in seconds) taken to build the indices on the various genomes.

## 3 Results

To assess the practical performance of SkipPatch, we measured index creation, update, and query times (for varying query lengths) as well as substring extraction times on four genomes: *E. coli* (4.6 MB), *D. melanogaster* (142.6 MB), *T. nigroviridis* (288.3 MB) and *H. sapiens* (2.98GB). These genomes were downloaded from NCBI's Assembly resource [1]. We preprocessed these genomes so that regions of the DNA sequence which are unknown are collapsed into a single *N* character and replaced with an *A*. This was done in order to simplify the procedure. If maintaining the offsets induced by the *N*s is important, one could simply leave the *N*s in the reference string but not hash any k-mers that consist of *N*s, since these k-mers are uninformative and no reads would map to such regions. However, the positions for the informative k-mers would remain consistent with the reference containing the *N*s. No repeats were masked in this step. All tests are performed on genomes processed in this way. We compared these timings to timings obtained with an existing Dynamic Extended Suffix Array (Dyn SA) implementation [14]. The sampling factor for both Dyn SA and SkipPatch implementation was set to 16 for *H. sapiens* and 1 for all the other genomes. This was done in order to reduce memory consumption. All of these tests were performed on a shared-user machine with 256 GB memory, and 4 hexacore Intel Xeon E5 processors (24 cores total) clocked at 2.6GHz. The SkipPatch implementation is single threaded, as is the implementation of Dyn SA.

### 3.1 Speed of index creation and updates

The initial creation of the index for SkipPatch is much faster than for Dyn SA (Table 1). Update operations are also considerably faster in SkipPatch than Dyn SA (Figure 2). The speed advantage of SkipPatch holds as the number of updates performed varies between 0.01%, 0.1%, 0.5%, 1%, and 5% of the size of the underlying genome. In SkipPatch, we expect the time taken to perform a series of updates to scale in the number of k-mers that need to be inserted or modified, as well as logarithmically in the number of edits performed so far. Empirically, we observe only a linear increase in the time taken to perform larger numbers of updates. One explanation for this behavior is that the logarithmic factor (which scales as the logarithm of the number of updates, not the size of the genome) remains very small, in which case operations such as k-mer lookup and hash table update dominate this logarithmic factor.

### 3.2 Speed of queries

SkipPatch is also faster than Dyn SA to query for strings of arbitrary lengths, even after a large number of edits have been made (Table 2). It is important to note that despite SkipPatch being an inverted lookup table built on fixed sized k-mers, it is capable of efficiently searching for query strings of variable lengths.

As the size of the skip list storing the edits grows, SkipPatch might take progressively longer to perform queries, as each query now has a term that is logarithmic in the number of edits performed so far. As with update speed, this extra logarithmic factor appears to have very little impact on query speed in practice, even when considering a substantially larger number of edits than one might expect in practice. In fact, as one considers a reference genome under which progressively more edits have been made, the query speed of SkipPatch remains very similar (Table 2).

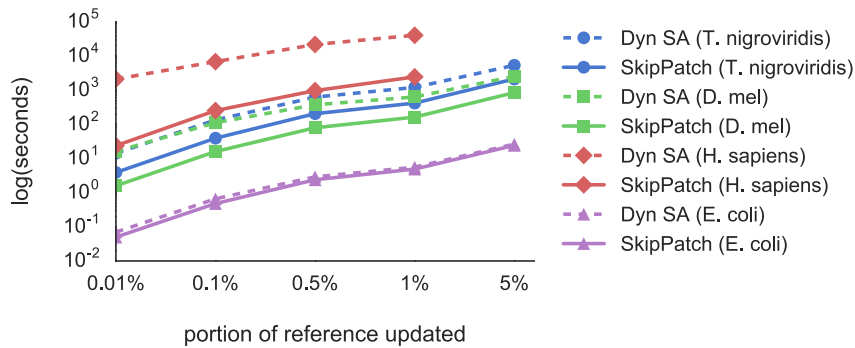

Figure 2: Time (in seconds) required for performing a number of updates equivalent to 0.01%, 0.1%, 0.5%, 1.0%, and 5.0%, of the underlying genome using both SkipPatch and Dyn SA. The average length of the updates is 5. The Dyn SA test for 5.0% updates in *H. sap* was aborted as it was running for over 24 hours as compared to SkipPatch, which took approximately 2.5 hours.

When testing queries on *T. nig* and *H. sap*, we noticed erratic search timings for SkipPatch— a small number of queries required orders of magnitude longer than the mean query time. We observed that these queries mapped to a small number of highly-repetitive regions. In order to deal with repetitiveness, prior work involving k-mer indices has adopted a strategy which involves eliminating k-mers corresponding to highly repetitive regions of the genome [10]. Thus, queries containing uninformative k-mers are filtered out. We define uninformative k-mers to be those which are among the top 0.01% most repetitive k-mers (corresponding to k-mers occurring more than 25 and 230 times for *T. Nig* and *H. Sap* respectively). We present the results for *T. nig* and *H. sap* for both filtered (repetitive k-mers removed) and non-filtered data (repetitive k-mers retained) (Table 2). The Dyn SA is faster when a query maps to highly repetitive regions. Repetitiveness in a genome did not effect the query times significantly for the Dyn SA. This could be due to the nature of the index, since it has to simply read off a list of positions once the query string is mapped to the appropriate interval on the suffix array. SkipPatch has a disadvantage in such a scenario because it has to consider all the positions to which every subread of the query string maps.

### 3.3 Substring extraction speed and memory usage

Using SkipPatch, one can extract an arbitrary substring substr($i, j$) between positions $i$ and $j$ in the edited string. Even after a number of updates have been performed, SkipPatch is able to do this much faster than Dyn SA (Table 3). This is likely partially due to the relative simplicity of the underlying operations required for extracting such a substring.

We also recorded the memory usage (before and after all of the updates have been performed) for both data structures (Figure 3). The memory consumed by SkipPatch can be attributed to the k-mer hash table, the genome itself, and the skip list. SkipPatch achieves faster speeds than the Dyn SA for updates, substring extraction and non-repetitive querying, and uses only marginally more memory than Dyn SA in most cases.

| Species | Method | 0.1% | 0.2% | 0.3% | 0.4% | 0.5% | 0.6% | 0.7% | 0.8% | 0.9% | 1% |
|---|---|---|---|---|---|---|---|---|---|---|---|
| *E. coli* ($l=40$) | Dyn SA | 0.23 | 0.23 | 0.23 | 0.23 | 0.23 | 0.22 | 0.22 | 0.22 | 0.22 | 0.22 |
| | SkipPatch | 0.15 | 0.14 | 0.14 | 0.15 | 0.14 | 0.14 | 0.14 | 0.15 | 0.14 | 0.14 |
| *E. coli* ($l=75$) | Dyn SA | 0.44 | 0.44 | 0.44 | 0.44 | 0.44 | 0.44 | 0.45 | 0.45 | 0.44 | 0.44 |
| | SkipPatch | 0.28 | 0.28 | 0.28 | 0.27 | 0.27 | 0.26 | 0.26 | 0.25 | 0.25 | 0.36 |
| *D. mel* ($l=40$) | Dyn SA | 0.71 | 0.69 | 0.79 | 0.85 | 0.90 | 0.95 | 0.99 | 0.99 | 0.97 | 1.00 |
| | SkipPatch | 0.60 | 0.49 | 0.51 | 0.54 | 0.41 | 0.51 | 0.51 | 0.42 | 0.55 | 0.43 |
| *D. mel* ($l=75$) | Dyn SA | 1.28 | 1.22 | 1.23 | 1.31 | 1.28 | 1.33 | 1.08 | 1.19 | 1.09 | 1.19 |
| | SkipPatch | 1.10 | 0.96 | 1.09 | 1.02 | 1.09 | 1.13 | 1.12 | 0.86 | 0.94 | 0.96 |
| *T. nig* ($l=40$) | Dyn SA | 1.51 | 1.33 | 1.49 | 1.92 | 1.74 | 2.19 | 1.32 | 2.21 | 2.02 | 1.14 |
| | SkipPatch | 19.62 | 15.33 | 15.02 | 16.18 | 18.10 | 17.08 | 14.33 | 20.58 | 15.08 | 9.20 |
| | Dyn SA (filtered) | 1.04 | 1.49 | 0.74 | 1.09 | 0.94 | 0.87 | 0.89 | 0.85 | 0.75 | 0.84 |
| | SkipPatch (filtered) | 0.19 | 0.19 | 0.21 | 0.20 | 0.20 | 0.20 | 0.21 | 0.19 | 0.20 | 0.20 |
| *T. nig* ($l=75$) | Dyn SA | 2.99 | 2.11 | 1.72 | 1.77 | 1.80 | 1.87 | 2.06 | 1.64 | 1.65 | 1.71 |
| | SkipPatch | 36.85 | 39.61 | 31.64 | 42.65 | 37.74 | 36.16 | 40.56 | 32.46 | 33.65 | 36.10 |
| | Dyn SA (filtered) | 1.13 | 1.22 | 1.20 | 1.20 | 1.20 | 1.33 | 1.22 | 1.22 | 1.22 | 1.31 |
| | SkipPatch (filtered) | 0.35 | 0.35 | 0.37 | 0.38 | 0.38 | 0.37 | 0.39 | 0.36 | 0.37 | 0.34 |
| *H. sap* ($l=40$) | Dyn SA | 19.48 | 13.22 | 13.67 | 18.39 | 14.92 | 18.16 | 14.15 | 18.92 | 14.07 | 17.73 |
| | SkipPatch | 40.52 | 45.33 | 34.04 | 38.25 | 3.67 | 47.09 | 68.82 | 50.90 | 45.27 | 65.83 |
| | Dyn SA (filtered) | 2.95 | 4.35 | 3.00 | 3.52 | 3.69 | 4.11 | 2.46 | 3.66 | 2.46 | 3.94 |
| | SkipPatch (filtered) | 0.71 | 0.79 | 0.85 | 0.73 | 1.19 | 1.03 | 0.85 | 1.07 | 0.99 | 1.03 |
| *H. sap* ($l=75$) | Dyn SA | 5.15 | 4.85 | 5.65 | 5.64 | 5.67 | 5.31 | 5.07 | 5.04 | 5.34 | 5.67 |
| | SkipPatch | 40.73 | 67.28 | 49.73 | 57.59 | 58.53 | 54.56 | 57.67 | 53.60 | 58.17 | 62.87 |
| | Dyn SA (filtered) | 4.25 | 4.59 | 4.40 | 6.75 | 5.88 | 5.56 | 4.57 | 5.32 | 4.70 | 5.40 |
| | SkipPatch (filtered) | 1.36 | 1.10 | 1.20 | 1.53 | 2.30 | 1.36 | 1.41 | 1.52 | 1.73 | 1.92 |

Table 2: Query time (in seconds) for 5000 strings of length 40 and 75 versus number of updates for 4 species. Number of updates are given as percentage of the underlying genome size. The results for *T. nig* and *H. sap* have been presented with and without filtered queries (details in section 3.2).

| Method | *E. coli* | *D. mel* | *T. nig* | *H. sap* |
|---|---|---|---|---|
| Dyn SA | 0.61 | 1.41 | 1.59 | 3.32 |
| SkipPatch | 0.03 | 0.04 | 0.05 | 0.08 |

Table 3: Time (in seconds) required to extract 10,000 random substrings of varying lengths after performing updates equal to 0.1% of the genome size. Updates were drawn from a Gaussian distribution with a mean length $\mu = 35$ and a standard deviation of $\mu/2$.

## 4 Discussion and Conclusion

We have introduced SkipPatch, a novel data structure for maintaining a hash-based index over a query string that is being dynamically updated. Our data structure is practically fast. Given the relatively small cost of performing updates and that querying the updated index is nearly as fast as querying the original hash, we believe that SkipPatch is a promising candidate around which to build a dynamic read mapping tool. This would allow for variant detection with the accuracy of iterative approaches [7], but with greatly improved speed.

While the dynamic suffix array has some very appealing properties, in practice, a number of factors have prevented it from being used to such an end. First is the theoretical issue that the complexity of an update (even a relatively small one) can, in the worst case, result in an update that is as expensive as re-building the entire data structure. We achieve a substantially improved worst-case update cost, which scales with the size of the update and the logarithm of the number of updates performed so far. Second, the practical lack of a high-quality implementation may have, to some degree, stymied the adoption of the dynamic suffix array. Though the implementation of SkipPatch that we provide is currently a prototype, we are continuing to work on improving the quality, speed and robustness of our implementation.

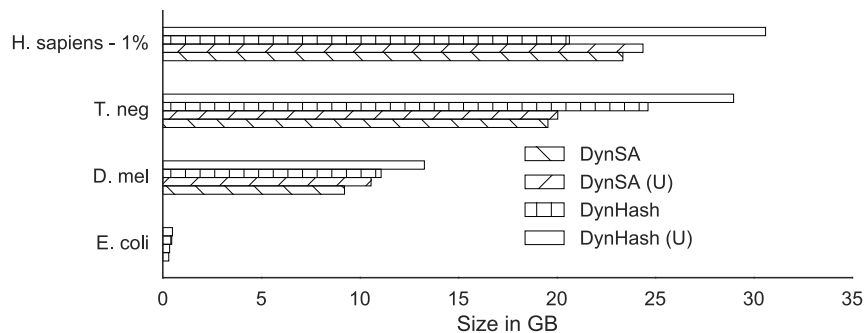

Figure 3: Memory (in GB) before and after performing updates equivalent to 5% of the underlying genome size (1% and a sampling factor of 16 in case of *H. sap* for both data structures). The average length of the updates is 5. *U* indicates the memory consumption after making the updates.

The design of SkipPatch has some additional useful properties when applied to the problem of read mapping (e.g. approximate pattern search for strings substantially longer than the length of the indexed substrings). Specifically, when using k-mer lookup to assist in read mapping, the k-mers that constitute the seeds for mapping a read are not independent. These seeds, since they originate from the same sequenced fragment, must occur nearby in the underlying target genome to be indicative of a high-quality mapping location. This suggests that one can make use of a batch query when applying SkipPatch to read mapping: rather than adjust the positions of each of the k-mers that are being queried individually, one can query the skip list for the k-mers that appear early on in the read to find an updated position nearby where the remaining k-mers must map if the retrieved location is a promising candidate. Such strategies further mitigate the already small cost of the extra logarithmic factor incurred by SkipPatch when performing queries.

We believe that SkipPatch represents a promising new direction in the efficient indexing of texts undergoing dynamic modification. Since it is not a *full-text* index, fundamentally different strategies for modifying the underlying text are applicable, leading to favorable asymptotic bounds. In addition to the theoretical benefits, we find that SkipPatch is fast in practice and can likely be adopted into higher-level tools where such dynamically updateable indices are necessary.

A prototype implementation of SkipPatch, written in `C++11` is available from `https://github.com/COMBINE-lab/SkipPatch` under the GPL v3 license.

## 5   Acknowledgments

We would like to thank Srikant Aggarwal, Hirak Sarkar, and Ayush Sengupta for useful ideas discussions regarding the SkipPatch data structure and algorithms. Tim Wall was a predoctoral trainee supported by NIH T32 training grant T32 EB009403 as part of the HHMI-NIBIB Interfaces Initiative. This research is funded in part by the Gordon and Betty Moore Foundation's Data-Driven Discovery Initiative through Grant GBMF4554 to Carl Kingsford, by the US NSF (1256087, 1319998), and by the US NIH (HG007104).